\documentclass[aps,prl,reprint,longbibliography,floatfix,amsmath,amssymb,showpacs,superscriptaddress,groupedaddress]{revtex4-1}

\usepackage{url,graphicx,tabularx,array,amsmath,listings,amssymb,float,dcolumn,bm,epsfig,epstopdf} 

\begin{document}
\renewcommand{\arraystretch}{1.8}

\title{Strongly Anisotropic Spin Relaxation in the Neutral Silicon Vacancy Center in Diamond}

\author{B. C. Rose$^1$}
\author{G. Thiering$^{2,4}$}
\author{A. M. Tyryshkin$^1$}
\author{A. M. Edmonds$^3$}
\author{M. L. Markham$^3$}
\author{A. Gali$^{2,4}$}
\author{S. A. Lyon$^1$}
\author{N. P. de Leon$^1$}
\email{npdeleon@princeton.edu}

\affiliation{$^1$Dept.\ of Electrical Engineering, Princeton University, Princeton, NJ 08544, USA}
\affiliation{$^2$Wigner Research Centre for Physics, Hungarian Academy of Sciences, PO Box 49, H-1525, Budapest, Hungary}
\affiliation{$^3$Element Six, Harwell, OX11 0QR, United Kingdom}
\affiliation{$^4$Department of Atomic Physics, Budapest University of Technology and Economics, Budafoki \'{u}t 8., H-1111 Budapest, Hungary}



\date{\today}

\begin{abstract}
Color centers in diamond are a promising platform for quantum technologies, and understanding their interactions with the environment is crucial for these applications. We report a study of spin-lattice relaxation ($T_1$) of the neutral charge state of the silicon vacancy center in diamond.  Above 20 K, $T_1$ decreases rapidly with a temperature dependence characteristic of an Orbach process, and is strongly anisotropic with respect to magnetic field orientation.  As the angle of the magnetic field is rotated relative to the symmetry axis of the defect, $T_1$ is reduced by over three orders of magnitude. The electron spin coherence time ($T_2$) follows the same temperature dependence but is drastically shorter than $T_1$.  We propose that these observations result from phonon-mediated transitions to a low lying excited state that are spin conserving when the magnetic field is aligned with the defect axis, and we discuss likely candidates for this excited state.
\end{abstract}

\maketitle
Solid state defects are attractive candidates for quantum technologies because they can have long spin coherence time and can be integrated into nanofabricated devices.  However, interactions with phonons in the solid state environment can lead to spin decoherence.  Color centers in diamond with exceptionally long spin coherence time have been identified, such as the nitrogen-vacancy center (NV$^-$), and these are promising candidates for a wide range of applications including quantum sensing \cite{Balasubramanian2008}, quantum information processing \cite{Gaebel2006,Dutt2007,Maurer2012,Dolde2013}, and quantum networks \cite{Togan2010,Pfaff2014,Hensen2015}. More recently the negatively-charged silicon vacancy center in diamond has been shown to possess promising optical coherence, but poor spin coherence at 4 K because of a phonon-mediated orbital relaxation process \cite{Jahnke2015,2016Sipahigil}. We recently demonstrated that the neutral charge state of the silicon vacancy color center (SiV$^0$) has excellent optical coherence, as well as long spin coherence times at temperatures up to 20 K \cite{Rose2017}.  These properties make it an ideal candidate for a single atom quantum memory in a quantum network.  However, we also observed that at temperatures above 20 K both $T_1$ and $T_2$ decrease exponentially with temperature.  Understanding the origin of this process is crucial for extending the operation range of SiV$^0$ to higher temperatures, to enable new applications in quantum information processing and nanoscale sensing.

\begin{figure}
\includegraphics[width=\linewidth]{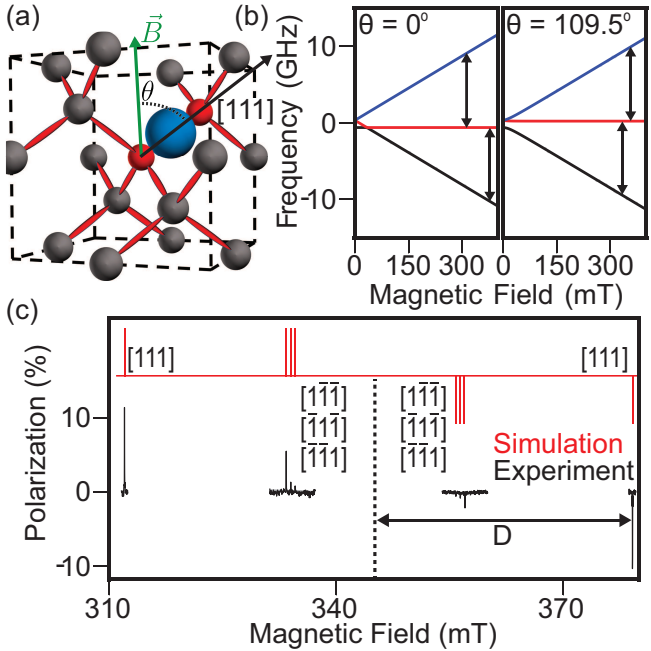}
\caption{(a) Ball and stick model of the silicon vacancy center in diamond. Gray spheres are carbon atoms. The interstitial Si atom (blue sphere) and split vacancy (red spheres) are aligned along the $\langle 111 \rangle$ directions in the diamond lattice, and the magnetic field (\textbf{B}) forms angle $\theta$ with the defect axis.  (b) Energies of the three ground state spin sublevels for the two inequivalent orientations of SiV$^0$ with $\textbf{B}\parallel \left[ 111 \right]$.  (c) Pulsed ESR spectrum of SiV$^0$ (zero field splitting, $D = 0.94$ GHz or 33.5 mT) measured at X-band frequency (9.7 GHz) with the magnetic field slightly misaligned from $\left[ 111\right]$ by $\theta=2.6^\circ$ (black), along with the simulated transitions according to Eq. \ref{eq:spinH} (red).  The four sets of lines correspond to the two transitions $m_s=0\leftrightarrow +1$ and $m_s=-1\leftrightarrow0$ for two inequivalent orientations. The simulation does not account for the difference in optical polarization between the sites.}
\label{fig:ESRIntro} 
\end{figure}

In this letter we investigate the spin-lattice relaxation of SiV$^0$ in detail.  The exponential temperature dependence of $T_1$ above 20 K is consistent with an Orbach process \cite{Orbach1961,Orbach1972Book,1983Shrivastava} with an activation energy ($E_a$) of 16.8 meV, and we observe that the relaxation rate has a sharp dependence on the angle ($\theta$) of the magnetic field ($\textbf{B}$) relative to the symmetry axis of the defect (Fig. \ref{fig:ESRIntro}a).  As the angle of the magnetic field is rotated away from the crystallographic axis of SiV$^0$ by just 5 degrees, $T_1$ decreases by almost two orders of magnitude.  In addition to the unusual orientation dependence of $T_1$, $T_2$ follows the same temperature dependence as $T_1$ but at a rate that is three orders of magnitude faster when the magnetic field is aligned with the defect symmetry axis.  

We propose that the strong intrinsic anisotropy in the spin-lattice relaxation of SiV$^0$ and the significantly shorter spin coherence time originate from the presence of phonon-mediated transitions to an excited state that are spin-conserving when the magnetic field is aligned with the quantization axis of the center. This is fundamentally similar to previous observations in SiV$^-$ at 4 K, in which a fast orbital relaxation ($T_{1,orbital}$ = 38 ns) is spin-conserving but spin-dephasing, giving rise to a relatively long $T_{1,spin}$ = 2 ms, while $T_2$ is limited by the orbital relaxation rate \cite{Rogers2014Optical,Jahnke2015}.  For SiV$^-$, spin relaxation arises from differing spin-orbit coupling in the two low-lying orbital states, and when the external magnetic field is aligned with the SiV$^-$ axis, this spin relaxation is suppressed, leading to $T_{1,spin}\gg T_{1,orbital}$. However, in a large off-axis magnetic field, the eigenstates mix, and the spin relaxation rate increases rapidly with angle \cite{Rogers2014Optical}.

In SiV$^0$, the identity of the low-lying excited state at 16.8 meV implied by the Orbach activation energy is unknown. Likely candidates are a low lying singlet state or a triplet vibronic mode \cite{Zaitsev2000}. For both candidates, we propose models for the suppression of spin relaxation when the external magnetic field is aligned with the quantization axis, which capture the temperature and orientation dependence of $T_1$ and the orientation-dependent ratio between $T_1$ and  $T_2$. We show that a model incorporating a singlet excited state closely reproduces our data, and we outline the physical requirements for a triplet excited state that would account for the experimental observations.

Two high purity $\left\{110\right\}$ diamonds grown by chemical vapor deposition were used in these experiments.  The first diamond (D1) was doped during growth with both boron ($\gtrsim10^{17}$ cm$^{-3}$) and silicon ($\sim 10^{17}$ cm$^{-3}$) and subsequently HPHT annealed, resulting in a SiV$^0$ concentration of $4\cdot10^{16}$ cm$^{-3}$ \cite{Edmonds2008}.  The silicon precursor was isotopically enriched with $90\%$ $^{29}$Si, and all measurements in D1 were conducted on a $^{29}$Si hyperfine line.  The second diamond (D2) was doped during growth with boron ($\sim10^{17}$ cm$^{-3}$) and implanted with $^{28}$Si ($6.3\cdot10^{15}$ cm$^{-3}$), and was previously described and characterized in reference \cite{Rose2017}.  After Si ion implantation and high temperature annealing, the resulting SiV$^0$ concentration was $5.1\cdot10^{15}$ cm$^{-3}$ within the implanted region.  Pulsed X-band (9.7 GHz) electron spin resonance (ESR) was performed in a standard dielectric volume resonator (Bruker MD5) with a quality factor of $Q\approx 5000$ \cite{1964Rosenbaum}.  The experimental apparatus is described in detail in reference \cite{Rose2017}. In all experiments the microwave power was chosen so that the excitation pulse bandwidth was greater than the bulk linewidth of spin transitions in all experiments ($\sim1$~MHz).  Measurements of $T_{2}$ utilized a standard two-pulse Hahn echo sequence with an initial 100 ms pulse of $\sim200$ mW of green laser light (532 nm) to optically enhance the spin polarization.  Under these conditions at 5 K, we achieve $11.5\%$ optical spin polarization into m$_s=0$ (Fig. \ref{fig:ESRIntro}c). $T_{1}$ was measured using a three-pulse inversion recovery sequence \cite{SchweigerBook}.

\begin{figure}
\includegraphics[width=\linewidth]{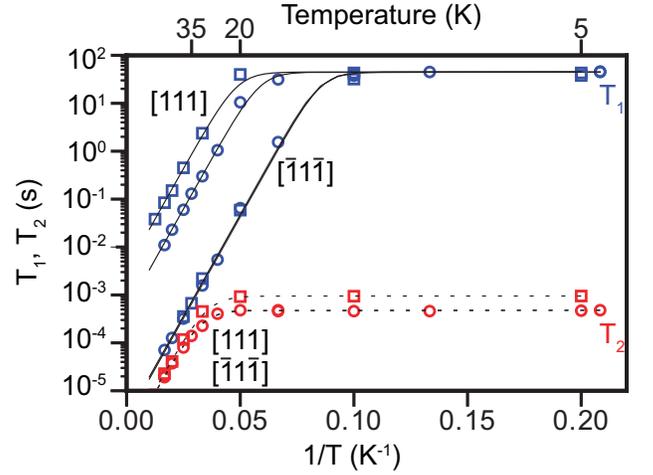}
\caption{Arrhenius plot of the temperature dependence of $T_1$ (blue) and $T_2$ (red) for SiV$^0$ in diamonds D1 (circles) and D2 (squares) for two inequivalent orientations with $\bm{B}\parallel\left[ 1 1 1\right]$.  $\left[111\right]$: $\theta=2.6^\circ$ (D1), $\theta=0.8^\circ$ (D2) and $\left[\bar{1}1\bar{1}\right]$: $\theta=106.9^\circ$ (D1), $\theta=108.7^\circ$ (D2).  The lines correspond to the best fit of Eq. \ref{eq:fitExpTdep}.}
\label{fig:T1T2Temp} 
\end{figure}

The ground state electron spin Hamiltonian of SiV$^0$ is given by \cite{Edmonds2008}:
\begin{equation}
\hat{H}=\hat{\bm{S}}^\dagger\tilde{D}_g\hat{\bm{S}} + \mu_B \hat{\bm{S}}^\dagger\tilde{g}\bm{B}\text{,}
\label{eq:spinH}
\end{equation}
with electron spin $S=1$, zero field splitting tensor ($\tilde{D}_g$) with axial part $D_g=0.94$ GHz (at T = 4.8 K), electron $g$ tensor ($\tilde{g}$) with parallel and perpendicular components $g_\parallel=2.0042$ and $g_\perp=2.0035$ respectively, and the Bohr magneton, $\mu_B$.  The $\tilde{D}$ and $\tilde{g}$ tensors are both aligned along the $\langle 111 \rangle$ directions.  With the field aligned along $\left[ 111\right]$, there are two inequivalent orientations (Fig. \ref{fig:ESRIntro}c):  one orientation aligned with the magnetic field so that $\theta=0^\circ$ ($\left[ 1 1 1 \right]$, outer ESR peaks), and three equivalent orientations aligned off-axis with $\theta=109.5^\circ$ ($\left[ \bar{1}\bar{1}1\right]$, $\left[ 1\bar{1}\bar{1}\right]$,$\left[ \bar{1}1\bar{1}\right]$, inner ESR peaks).  

We performed time-resolved measurements for both inequivalent orientations to study spin relaxation and decoherence.  Below 20 K, $T_1$ and $T_2$ are independent of temperature (Fig. \ref{fig:T1T2Temp}).  In sample D2, we previously reported that SiV$^0$ exhibits a spin coherence time at low temperature that is dominated by spectral diffusion from the 1.1$\%$ abundance of $^{13}$C nuclei, with $T_{2}=0.954\pm 0.025$ ms \cite{Rose2017}.  By contrast, the density of SiV$^0$ in sample D1 is large enough that the spin coherence time is limited by instantaneous diffusion, with $T_{2}=0.48\pm 0.03$ ms \cite{Tyryshkin2012, Supplemental}.  At low temperature, $T_1$ is independent of temperature for both samples, with $T_{1}=46\pm2$~s (D1) and $T_{1}=45\pm4$~s (D2).  This saturation of $T_1$ at low temperature is similar to previous observations of NV$^-$ \cite{2012Jarmola}.

Above 20 K, both $T_{1}$ and $T_{2}$ decrease exponentially with increasing temperature. In this high temperature regime the two inequivalent orientations ($\left[ 111\right]$ and $\left[ \bar{1}1\bar{1}\right]$) exhibit similar $T_2$ but significantly different $T_1$.  $T_1$ and $T_2$ exhibit the same Arrhenius slope for both orientations.  The data ($T_{1,\left[111 \right]}$, $T_{1,\left[ \bar{1}1\bar{1} \right]}$, $T_{2,\left[ 111 \right]}$, $T_{2,\left[ \bar{1}1\bar{1} \right]}$) were fit according to the equation:
\begin{equation}
\frac{1}{T_{1,2}} = \frac{1}{T_{sat}}+A\left( \theta \right) e^{-E_a/kT}\text{,}
\label{eq:fitExpTdep}
\end{equation}
where $T_{sat}$ is the saturated decay time at low temperature, $A\left(\theta\right)$ is the orientation-dependent rate prefactor, $E_a$ is the activation energy, and $kT$ is the thermal energy. The activation energy is the same for all curves, $E_a=16.8\pm1.5$ meV, but $A\left( \theta \right)$ varies significantly (Table \ref{tab:fitParam}).  

Unlike $T_1$, $T_2$ exhibits a weak orientation dependence above 20 K ($T_{2,\left[111 \right]}\approx T_{2,\left[ \bar{1}1\bar{1} \right]}$).  However, since $T_2$ displays the same activation energy as $T_1$, the two decay times likely result from the same physical process.  This is surprising since $T_2$ is not $T_1$-limited; in fact $T_2$ is 4000 times shorter than $T_1$ when $\theta=0^\circ$.  We can rule out that the decoherence is caused by magnetic noise from nearby centers with short $T_1$ because we do not observe a density dependence in $T_2$  when comparing samples D1 and D2, and we are unable to extend $T_2$ with further dynamical decoupling \cite{Supplemental}\cite{Rose2017}.  Moreover, numerical simulations of ensemble dipolar interactions fail to account for the observed temperature dependence of $T_2$ \cite{Supplemental}.

\begin{table}
\begin{ruledtabular}
\caption{Summary of the rate prefactors, $A\left(\theta\right)$, extracted from the curves in Fig. \ref{fig:T1T2Temp} using Eq. \ref{eq:fitExpTdep}.}
\begin{tabular}{cccc}
  Sample && A$\left(\theta\right)$ (kHz)& \\\cline{2-4}
  &$T_1$, $\left[111\right]$&$T_1$, $\left[\bar{1}1\bar{1}\right]$&$T_2$, all orientations\\
  \hline
  D1&$2.10\pm 0.28$&$378\pm33$&$1260\pm152$\\
  D2&$0.3\pm 0.02$&$365\pm53$&$1180\pm210$\\
\end{tabular}
\label{tab:fitParam}
\end{ruledtabular}
\end{table}

\begin{figure}
\includegraphics[width=\linewidth]{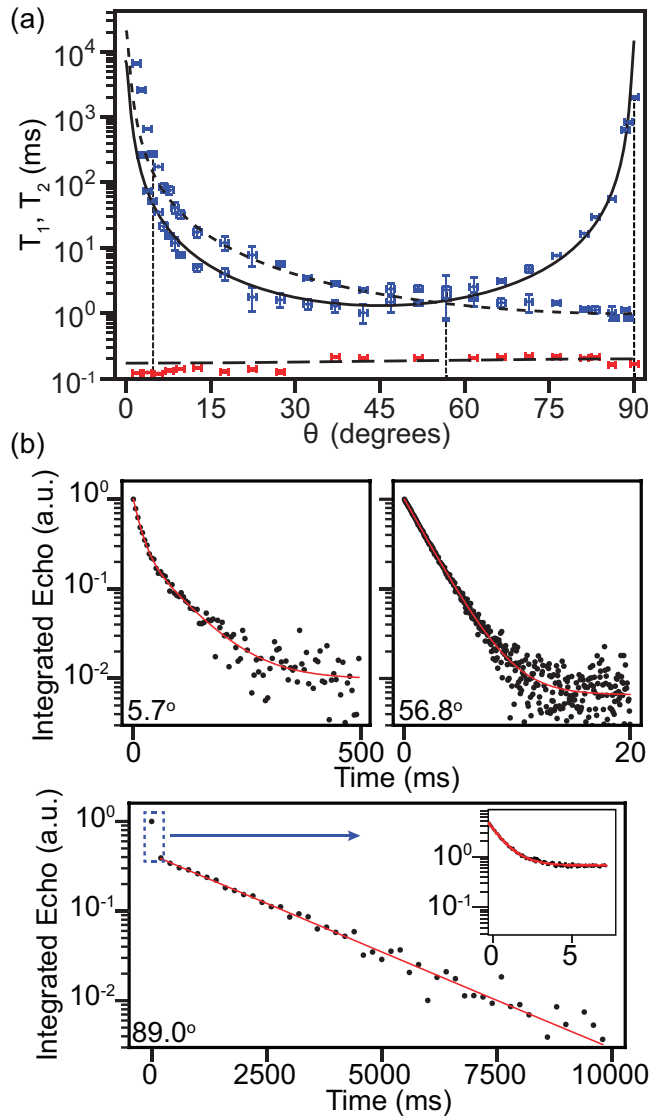}
\caption{(a) Orientation dependence of $T_1$ (blue dots) and $T_2$ (red dots) in sample D1 at $T = 30$ K, measured on the m$_s=0\leftrightarrow+1$ transition.  Lines show the theoretical orientation dependence of the two characteristic spin relaxation times $T_{1,a}$ (solid line) and $T_{1,b}$ (dashed line), as well as $T_2$ (long dashed line) predicted for an Orbach process with an excited singlet state.  (b) Selected decay curves with their corresponding fits (red) showing the biexponential behavior of the spin relaxation at particular magnetic field orientations ($\theta$), indicated by the vertical dashed lines in panel (a).}
\label{fig:T1Orientation} 
\end{figure}

In order to understand the anisotropy in detail we measured the full orientation dependence of $T_1$ and $T_2$ at 30~K (Fig. \ref{fig:T1Orientation}), where the Orbach process dominates the spin relaxation. At this magnetic field ($\sim 3400$ G, Fig. \ref{fig:ESRIntro}b) the Zeeman frequency ($9.7~\text{GHz}$) is much larger than the zero field splitting ($0.94~\text{GHz}$).  The relative orientation of the magnetic field was varied by rotating the crystal about a $\langle 110 \rangle$ axis from $\theta=0^\circ$ ($\bm{B}\parallel\bm{D}$) to $\theta=90^\circ$ ($\bm{B}\perp\bm{D}$). The ESR spectrum (Fig. \ref{fig:ESRIntro}c) was measured to determine the crystal orientation to within $1^\circ$. The relaxation time exhibits dramatic anisotropy, and as the crystal is rotated away from $\theta=0^\circ$, the spin relaxation becomes clearly biexponential (Fig. \ref{fig:T1Orientation}). Near $\theta=0^\circ$, $T_1$ drops rapidly, and rotating by just $5^\circ$ increases the relaxation rate by almost two orders of magnitude. Near $\theta = 55^\circ$, the decay is a single exponential with a short timescale that is insensitive to small rotations. Beyond $55^\circ$ the two timescales diverge and differ by over 3 orders of magnitude at $\theta=90^\circ$ (Fig. \ref{fig:T1Orientation}b).

\begin{figure}
\includegraphics[width=\linewidth]{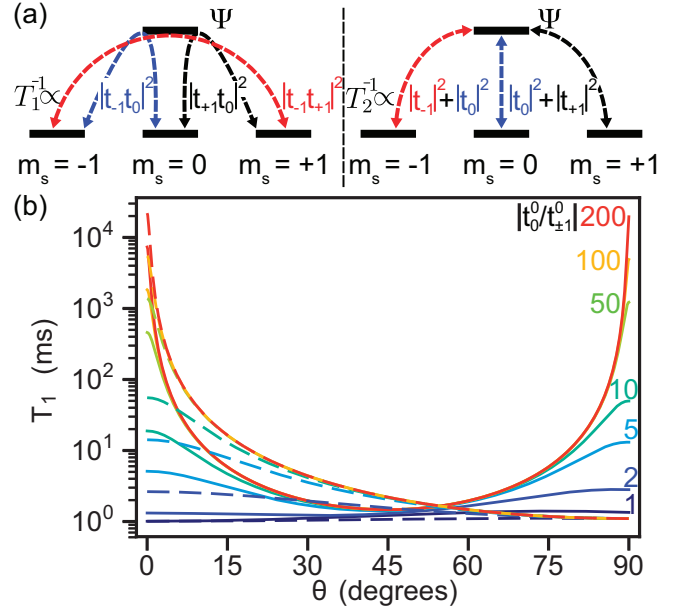}
\caption{(a) Level diagram of the Orbach process with a singlet excited state. Spin relaxation (left) occurs in two steps through the excited state and depends on the product of the overlap parameters, while decoherence (right) can arise from a single step, and depends on the sum.  (b) Plot of $T_{1,a}$ (solid curves) and $T_{1,b}$ (dashed curves) for selected values of $\left| t_0^0 / t_{\pm1}^0 \right|$ using the rate coefficient C extracted from experiment.}
\label{fig:PJTEModel} 
\end{figure}

We propose a model that captures the four salient features of the data: (1) the strong anisotropy of $T_1$, (2) the biexponential nature of $T_1$, (3) the temperature dependence of $T_2$, and (4) the large ratio between $T_1$ and $T_2$.  Generically, an Orbach process is a two-phonon relaxation process \cite{1983Shrivastava} that connects the ground state spin sublevels $m_s=-1,0,+1$ through a low-lying excited state ($\mid \!\!\Psi \rangle$) with amplitudes $t_{-1}$, $t_0$, and $t_{+1}$, respectively (Fig. \ref{fig:PJTEModel}a). The amplitudes ($t_{-1}$, $t_0$, and $t_{+1}$) are overlap parameters between the ground triplet states and the excited state, $t_{m_s}=\langle m_s\!\!\mid\!\! \Psi\rangle$ \cite{Supplemental}.  This gives rise to three possible relaxation rates between distinct pairs of the ground state triplet spin sublevels $m_s \leftrightarrow  m_{s'}=-1\leftrightarrow 0,0\leftrightarrow+1,-1\leftrightarrow +1$:
\begin{equation}
\frac{1}{T_{1,m_s\leftrightarrow m_{s'}}}=C \left| t_{m_s} t_{m_{s'}} \right|^2e^{-E_a/kT}\text{,}
\label{eq:T1General}
\end{equation}
where $C$ is a constant.  

If the excited state $\Psi$ is a singlet state ($S=0$), it is invariant under magnetic field orientation, so the behavior of T$_1$ can be captured by considering the mixing of the ground state \cite{Supplemental}.  The mixing of the spin sublevels in the presence of a large off-axis magnetic field leads to:

\begin{equation}
\begin{pmatrix}
   \left|t_{-1}\right|^2\\
   \left|t_{0}\right|^2\\
   \left|t_{+1}\right|^2
\end{pmatrix}
=
\begin{pmatrix}
   \cos^4{\frac{\theta}{2}} & \frac{1}{2}\sin^2{\theta} & \sin^4{\frac{\theta}{2}}\\
   \frac{1}{2}\sin^2{\theta} & \cos^2{\theta} & \frac{1}{2}\sin^2{\theta}\\
   \sin^4{\frac{\theta}{2}} & \frac{1}{2}\sin^2{\theta} & \cos^4{\frac{\theta}{2}}
\end{pmatrix}
\begin{pmatrix}
    \left|t_{-1}^0\right|^2\\
    \left|t_{0}^0\right|^2\\
    \left|t_{+1}^0\right|^2
\end{pmatrix}\text{,}
\label{eq:t_angle}
\end{equation}
where $\left|t_{m_s}^0\right|^2$  are the overlap parameters at zero magnetic field.  Substituting Eq. \ref{eq:t_angle} in Eq. \ref{eq:T1General} and solving the 3x3 relaxation rate matrix equation for the ground state spin (S=1) provides the $T_1$ relaxation times \cite{Supplemental}.  If $\left|t_0^0\right| = \left|t_{-1}^0\right|= \left|t_{+1}^0\right|$, then Eqs. \ref{eq:T1General} and \ref{eq:t_angle} predict that the spin relaxation is isotropic. However, if $\left|t_0^0\right| \gg \left|t_{-1}^0\right|, \left|t_{+1}^0\right|$, the spin relaxation is strongly anisotropic with two characteristic times approximated as:
\begin{equation}
\begin{split}
  &\frac{1}{T_{1,a}}=\frac{3}{8}C\left|t_{0}^0\right|^{4}\sin^2(2\theta)e^{-E_a/kT} \\
  &\frac{1}{T_{1,b}}=\frac{1}{2}C\left|t_{0}^0\right|^{4}\sin^2(\theta)e^{-E_a/kT}\text{.}
\end{split}
\label{eq:T1e_angle}
\end{equation} 
In this limit the model captures the observed angular dependence of the two timescales in $T_1$ as shown in Fig.~\ref{fig:T1Orientation}a.  By comparing numerical calculations of the orientation dependence of $T_1$ for different ratios of $\left|t_0^0/t_{\pm1}^0\right|$ (Fig. \ref{fig:PJTEModel}b), we can place a lower bound on the imbalance between these rates, $\left|t_0^0/t_{\pm1}^0\right| > 100$ \cite{Supplemental}. 

We can also predict the effect of this Orbach process on $T_2$. Customarily, the Orbach process is viewed as a spin relaxation process \cite{1983Shrivastava}. However, transitions to the excited state via absorption and emission of phonons can also lead to decoherence even when the spin projection is preserved, similar to what has been observed for orbital relaxation in SiV$^-$ \cite{Rogers2014Optical}. While the spin relaxation rate relies on a spin flip and therefore the product of the overlap parameters $\frac{1}{T_1}\propto \left| t_{m_s}\right|^2\left|t_{m_{s'}}\right|^2$ (Fig. \ref{fig:PJTEModel}a, left), the decoherence rate depends on the sum of overlap parameters, $\frac{1}{T_2}\propto \left| t_{m_s}\right|^2+\left|t_{m_{s'}}\right|^2$ (Fig \ref{fig:PJTEModel}a, right), if we assume the spin coherence is completely lost within a single cycle. The observed ratio of $T_1$ to $T_2$ will therefore depend on both the angle of the magnetic field and the ratio of overlap parameters. More accurately, the model predicts (Fig. \ref{fig:T1T2RatioOrientation}):
\begin{equation}
\frac{T_{1,a}}{T_{2,0\leftrightarrow \pm1}}=\frac{\left(\left| t_{\pm1}\right|^2+ \left|t_0\right|^2\right)\left( \left|t_0^0\right|^2+2\left|t_{\pm1}^0\right|^2\right)}{3\left| t_{\pm1} t_0 \right|^2}\text{.}
\label{eq:T1T2Rat}
\end{equation} 
The orientation dependence of $T_2$ predicted from this model is plotted in Fig. \ref{fig:T1Orientation}a, where we also included the effect of instantaneous diffusion in sample D1, and is plotted in detail in Fig. \ref{fig:T1AngT0Tm}.  The anisotropy in $T_2$ is mostly canceled in the sum $\left| t_{m_s}\right|^2+\left|t_{m_{s'}}\right|^2$.  The model provides the best fit for both the $T_1$ and $T_2$ data when $\left|t_0^0/t_{\pm1}^0\right|\approx 125$ (Fig. \ref{fig:T1Orientation}a).  

If the excited state $\Psi$ is instead a triplet state (S=1), then the overlap parameters cannot be written in a compact form, but we analyze this case in detail in the supplementary information \cite{Supplemental}. Briefly, phonon-mediated orbital relaxation to a vibronic excited state is generally spin conserving, but differences in the ground and excited state spin Hamiltonians can lead to mixing during the time spent in the excited state. Specifically, for SiV$^0$ the ground and excited states can have different zero field splitting tensors ($D_e$ and $D_g$). Since the Zeeman splitting in these measurements is 9.7 GHz, the zero field splittings must differ by a comparable scale in order to reproduce the observed ratio of $T_1$ to $T_2$, and we find that the data can be qualitatively reproduced when $D_e \sim 5-7$ GHz (Fig. \ref{fig:T2Orientation}b), compared to the ground state zero field splitting, $D_g = 0.94$ GHz. It is unlikely that the zero field splittings differ by such a large magnitude. Alternatively, the small ratio of $T_1$ to $T_2$ could also arise from incomplete spin dephasing. If the excited state lifetime is short compared to the spin precession time ($\tau < \pi\hbar/ E_{Zeeman} \sim 50$ ps), then the spin coherence is partially preserved in an excitation cycle \cite{VantHof1976}. A model involving a triplet excited state would therefore require either that $D_e \gg D_g$ or that the excited state lifetime is short enough to partially preserve coherence.

In summary, we have shown that spin relaxation in SiV$^0$ at high temperature is dominated by an Orbach process that is strongly dependent on the magnetic field orientation, and $T_2$ exhibits the same temperature dependence as $T_1$, but at a significantly faster rate. These observations can be explained by a model for the Orbach process where the overlap parameters from the $m_s=0$ and $m_s=\pm1$ spin sublevels to a singlet excited state are drastically different. We note that this imbalance in overlap parameters is consistent with the preferential optical spin polarization through the intersystem crossing into $m_s=0$ (Fig. \ref{fig:ESRIntro}c) \cite{Rose2017}. Alternatively, these observations can be qualitatively reproduced by a model with a triplet excited state that either exhibits a much larger zero field splitting than the ground state or a very short excited state lifetime. Although our present results cannot definitively identify the excited state, detailed spectroscopy can help distinguish between these two cases. For example, absorption spectroscopy of different isotopes could elucidate the vibronic structure \cite{Kehayias2013,Huxter2013}, and the nature of the singlet state can be explored using time-resolved photon correlation measurements, as well as temperature-dependent intersystem crossing rates \cite{Drabenstedt1999, Jelezko2006}. Furthermore, at temperatures well above the activation energy, there should be enough population in the excited state to observe spin resonance transitions associated with a spin-triplet state with different zero field splitting. We have not observed the existence of additional transitions with large zero field splitting, but on-going work includes increasing our measurement sensitivity at higher temperatures to search thoroughly for such states.

The strong intrinsic anisotropy in the spin-lattice relaxation of SiV$^0$ stands in contrast to prior studies of NV$^-$, in which spin relaxation is mostly insensitive to the magnetic field orientation (except in cases where the defect density is high enough that the relaxation is dominated by dipolar interactions \cite{Mrozek2015}). To the best of our knowledge, there has not been a detailed study of the orientation and temperature dependence of spin relaxation in NV$^-$ at high magnetic fields, and it would be interesting to perform such measurements in light of our work. Similarly, a more detailed orientation and temperature dependence of $T_{1,spin}$ in SiV$^-$ would further elucidate analogous spin and orbital relaxation processes, and recent measurements at dilution refrigerator temperatures have started to explore the mechanisms for spin relaxation and decoherence \cite{Becker2017,Sukachev2017}. 

Additionally, these observations point to a promising avenue of exploration for high temperature operation if the excited state involved in the Orbach process is a spin singlet state.  The imbalance of the overlap parameters to the excited state implies that superpositions of $m_s=\pm1$ could have longer spin coherence time than even the measured single-quantum $T_1$.  Future experiments include double quantum spin resonance measurements to interrogate the coherence of such superposition states.

\section{\label{sec:Acknowledgements} Acknowledgements}
This work was also supported by the Princeton Center for Complex Materials, a NSF Materials Research Science and Engineering Center (grant No. DMR-1420541) and the NSF EFRI ACQUIRE program (grant No. 1640959).  G. Thiering and A. Gali were supported by the ÚNKP-17-3-III New National Excellence Program of the Ministry of Human Capacities and the EU Commission (DIADEMS Project Contract No. 611143).  The authors gratefully acknowledge Jeff Thompson, Shimon Kolkowitz, and Ashok Ajoy for helpful discussions. The authors would also like to thank Sorawis Sangtawesin and Zihuai Zhang for help in proofreading the manuscript.

\bibliography{SiV0PJTE_references_brose} 

\clearpage
%

%
%
%
%


\setcounter{figure}{0} 
\renewcommand{\thefigure}{S\arabic{figure}}
\renewcommand{\theequation}{S\arabic{equation}}
\section{Strongly Anisotropic Spin Relaxation in the Neutral Silicon Vacancy Center in Diamond: Supplementary material}

%
%
%


%

\subsection{Instantaneous diffusion in sample D1}
At low temperatures, the coherence time in sample D1 is limited by instantaneous diffusion, which arises when a microwave pulse induces spin flips on a dense bath of paramagnetic centers. If we consider a central spin surrounded by its neighbors, then the pulse will induce rotations of the neighbors as well as the central spin. Phase resulting from the pulse-induced change in the dipolar magnetic field is not refocused during a Hahn echo sequence, limiting the coherence time to $T_{2(ID)}$.

\begin{figure}
\includegraphics[width=\linewidth]{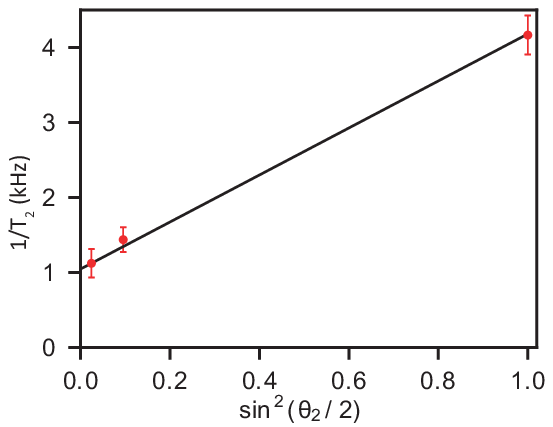}
\caption{Decoherence rates for SiV$^0$ centers in sample D1 measured at 5 K as a function of the rotation angle ($\theta_2$) of the second pulse in a Hahn echo sequence.  The linear dependence confirms that $T_2$ is limited by instantaneous diffusion.  The black curve is a fit according to Eq. \ref{eq:T2Extrap}.}
\label{fig:T2Extrapolation} 
\end{figure}

The effect of instantaneous diffusion can be mitigated by using a smaller rotation angle ($\theta_2$) for the second microwave pulse of a Hahn echo sequence, since the change in the net dipolar magnetic field scales as $\sin^2(\theta_2/2)$.  This results in a proportionally smaller phase accumulated by the central spin and a decoherence rate 1/$T_{2}\propto T_{2(ID)}^{-1} \sin^2(\theta_2)$.  Using a smaller rotating angle will enhance $T_2$, but it will also decrease the bulk echo signal by the same factor. In sample D1, the apparent decoherence rate increases linearly with $\sin^2(\theta_2/2)$ (Fig. \ref{fig:T2Extrapolation}).  The data were fit according to the following:

\begin{equation}
\frac{1}{T_2} = \frac{1}{T_{2(SD)}}+\frac{1}{T_{2(ID)}} \sin^2{\left(\theta_2/2\right)}\text{,}
\label{eq:T2Extrap}
\end{equation}
where $T_{2(SD)}$ is the spectral diffusion decay time.  The fit results in $T_{2(SD)}=0.95\pm0.22$ ms, most likely arising from the $1.1\%$ of $^{13}$C nuclei \cite{Rose2017} and $T_{2(ID)}=0.319\pm0.056$ ms. 

We note that the Hahn echo spin coherence times reported for sample D1 in Figs. \ref{fig:T2Extrapolation}, \ref{fig:STCompare}c, and \ref{fig:T2Orientation} ($T_2=0.28$ ms) is not the same as the spin coherence time reported in Fig. \ref{fig:T1T2Temp} in the main text ($T_2=0.48$ ms). This arises from the nonuniform population distribution of SiV$^0$ centers in this sample over the four inequivalent crystal orientations ($\left[111\right]$,$\left[1\bar{1}\bar{1}\right]$, $\left[\bar{1}1\bar{1}\right]$, $\left[\bar{1}\bar{1}1\right]$), which has been reported previously as sample C in reference \cite{EdmondsThesis2008}. The data in Figs. \ref{fig:T1Orientation}, \ref{fig:T1AngT0Tm}, \ref{fig:STCompare}, \ref{fig:T2Orientation}, and \ref{fig:T2Extrapolation} is taken using the $\left[\bar{1}1\bar{1}\right]$ orientation (smaller SiV$^0$ concentration), while the data in Fig. \ref{fig:T1T2Temp} is taken using the $\left[111\right]$ orientation (larger SiV$^0$ concentration).

\subsection{Decoherence arising from $T_1$-induced spin flips of fast relaxing neighbors}
An alternative hypothesis for the observed temperature dependence of $T_2$ (Fig. \ref{fig:T1T2Temp} in the main text) and its relative magnitude with respect to $T_1$ is that rapid dephasing arises from dipolar interactions with other SiV$^0$ spins in the bath, such as those misaligned with the external magnetic field. We can immediately rule out spectral diffusion from SiV$^0$ spin flip-flops and instantaneous diffusion mechanisms arising from dipolar interactions between SiV$^0$ centers since these mechanisms would be independent of temperature.  Instead, we consider the contribution of spectral diffusion arising from the fast $T_1$ relaxation of nearby SiV$^0$ centers \cite{Mims1968}. This decoherence mechanism is strongest when $T_1$ of the spin bath is comparable to the $T_2$ of the central spin under consideration.

\begin{figure}
\includegraphics[width=\linewidth]{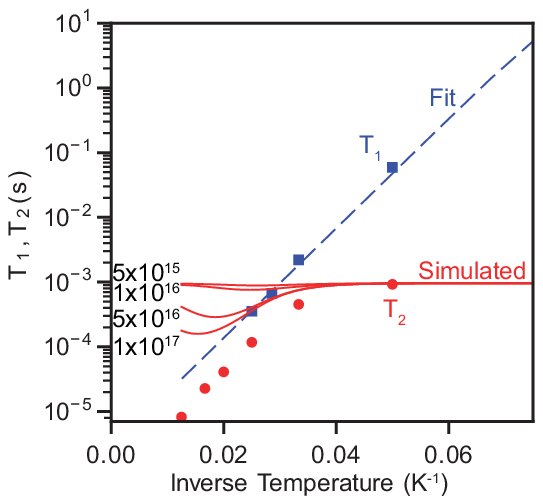}
\caption{Arrenhius plot of simulations of $T_2$ (red lines) resulting from spectral diffusion arising from fast relaxing SiV$^0$ centers. The blue dashed line is a fit of Eq. \ref{eq:fitExpTdep} to the temperature dependence of $T_1$ (blue squares) of fast relaxing SiV$^0$ sites. This fit is incorporated in Eq. \ref{eq:T1noiseT2} to simulate $T_2$ for a range of defect densities (labels in units of cm$^{-3}$).  The simulations indicate that for the range of densities studied, the decoherence arising from spin flips of nearby SiV$^0$ centers is not significant and is inconsistent with the observed temperature dependence of $T_2$ (red dots).}
\label{fig:T1NoiseSim} 
\end{figure}

In our samples, $T_2\sim0.5$ ms at low temperatures, which is comparable to $T_1\sim1$ ms of the three equivalent SiV$^0$ orientations misaligned with the magnetic field ($\theta\approx109^\circ$) at temperatures above 20 K.  We numerically model the contribution to the Hahn echo decay from these three equivalent off-axis sites for the range of densities in samples D1 and D2 \cite{Salikhov1981}.  The electron spin Hamiltonian describing a pair of SiV$^0$ spins, $S_1$ and $S_2$, is given by:

\begin{equation}
\begin{split}
\hat{H} =~&\hbar\omega_1 \hat{S}_{1z} + \hbar\omega_2 \hat{S}_{2z} + \hbar A\left(\bm{r}_{12}\right)\hat{S}_{1z}\hat{S}_{2z}\\
&+\hat{V}_2(t)\text{,}
\end{split}
\label{eq:pcH}
\end{equation}
with dipolar interaction between the SiV$^0$ spins
\begin{equation}
A\left(\bm{r}_{12}\right) = g_{1z} g_{2z} \mu_B^2 \hbar^{-1} \left( 1 - 3 \cos^2(\theta_{12})\right)r_{12}^{-3}\text{,}
\end{equation}
where $\omega_1$ and $\omega_2$ are the transition frequencies of the spins,  r$_{12}$ is the distance between the spins, $\theta_{12}$ is the angle between $\bm{r}_{12}$ and $\textbf{B}$, $g_{1z}$ and $g_{2z}$ are the longitudinal components of the $g$ tensors.  For our model we consider that $S_1$ is a slow relaxing spin ($\theta=0^\circ$) whose coherence time is being measured, and $S_2$ is a fast relaxing spin ($\theta=109^\circ$) whose spontaneous $T_1$ flips induce decoherence of $S_1$.  The term $\hat{V}_2(t)$ accounts for the fast Orbach spin relaxation rate of $S_2$ spins by inducing random spin flips at a rate $W$.  The contribution to the echo signal decay for $S_1$ is \cite{Salikhov1981}:

\begin{equation}
\begin{split}
V\left(2\tau\right) = &\left[ \left( \cosh(R\tau)+\frac{W}{R}\sinh(R\tau)\right)^2\right.\\
 &\left.+ \frac{A^2\left(\bm{r_{12}}\right)}{4 R^2}\sinh(R\tau)\right] \exp{\left(-2W\tau\right)}\text{,}
\end{split}
\label{eq:T1noiseT2}
\end{equation}
where $\tau$ is the inter-pulse delay in a Hahn echo sequence, $W=1/T_{1,\langle \bar{1}1\bar{1}\rangle}$ (fast relaxing sites, Fig. \ref{fig:T1NoiseSim} blue line), $R^2 = W^2-A^2\left(\bm{r_{12}}\right)/4$, and $r_{12}=n^{-1/3}$ is the average inter-spin distance.  This expression is averaged over all angles $\theta_{12}$ and added to the Hahn echo decay that arises from $^{13}$C spectral diffusion alone ($T_{2(SD)}=0.95$ ms).  The resulting calculated Hahn echo decay times are shown in Fig. \ref{fig:T1NoiseSim} for several SiV$^0$ densities in and above the range of the two samples studied here, which have SiV$^0$ concentrations of less than $5\cdot10^{16}$ cm$^{-3}$. The density required to account for the data would need to be 100 times higher. Furthermore, at high temperatures, motional narrowing should lead to an increase in $T_2$, which does not qualitatively agree with the observed temperature dependence (Figs. \ref{fig:T1T2Temp} and \ref{fig:T1NoiseSim}).

\subsection{Dynamical Decoupling using CPMG}
We previously reported dynamical decoupling measurements using the Carr-Purcell-Meiboom-Gill sequence on sample D2 \cite{Rose2017, Meiboom1958}. The Hahn echo $T_2$ displays a plateau below 20 K corresponding to $^{13}$C spectral diffusion, but is limited by an Orbach process above 20 K.  We observed that $T_{2,CPMG}$ is unchanged above 20 K and follows the temperature dependence of $T_2$.  However below 20 K $T_{2,CPMG}$ becomes substantially longer than $T_2$ and follows the extrapolated temperature dependence of the Orbach process.  We hypothesize that the CPMG experiment refocuses slow spectral diffusion that arises from the $^{13}$C nuclei, but it does not refocus fast effects from the Orbach process, as expected.  All of the points in the CPMG measurement lie along the same curve $T_{2,CPMG}^{-1}=A\exp{\left(-E_a/kT\right)}$, where $A=1180\pm210$ kHz and $E_a=16.8\pm1.5$ meV in the entire measured temperature range 5 K - 60 K. 

\subsection{Orientation dependence of $T_1$ and $T_2$ measured on $m_s=-1\leftrightarrow0$}
\begin{figure}
\includegraphics[width=\linewidth]{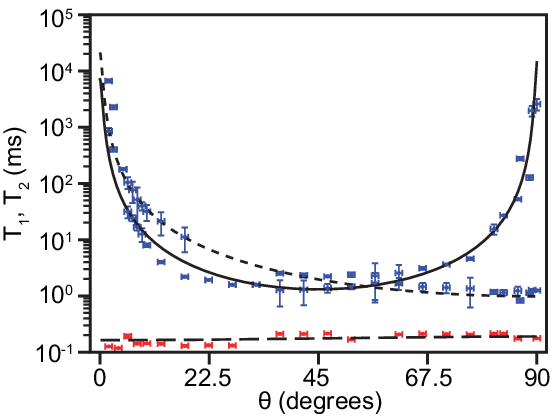}
\caption{Orientation dependence of $T_1$ (blue) and $T_2$ (red) measured on the $m_s=-1\leftrightarrow0$ transition.  The lines were simulated for the Orbach model with a singlet excited state using Eqns. \ref{eq:eigenRate} and assuming $\left|t_0^0/t_{\pm1}^0\right|=125$: (solid line) $T_{1,a}$, (dashed line) $T_{1,b}$, and (long dash) $T_2$.}
\label{fig:T1AngT0Tm} 
\end{figure}

In the main text we presented the orientation dependence of the $T_1$ and $T_2$ times for SiV$^0$ for measurements on the $m_s=0\leftrightarrow+1$ transition.  We also repeated the same measurements on the $m_s=-1\leftrightarrow0$ transition and find that it gives a nearly identical orientation dependence (Fig. \ref{fig:T1AngT0Tm}).  Since $T_{1,-1\leftrightarrow 0}\approx T_{1,0\leftrightarrow 1}$, we can conclude that $\left|t_{+1}^0\right|\approx\left|t_{-1}^0\right|$.  Additionally, because of the 1 GHz zero field splitting of SiV$^0$, the measurements on the $m_s=-1\leftrightarrow0$ transition were made at a field that was $\sim 300$ G larger (when aligned with the $\left[111\right]$ direction) compared to the measurements on the $m_s=0\leftrightarrow+1$ transition in Fig. \ref{fig:T1Orientation}.  This implies that the Orbach process has a weak dependence on the magnetic field strength.

\subsection{Ratio of $T_1$ to $T_2$}

\begin{figure}
\includegraphics[width=\linewidth]{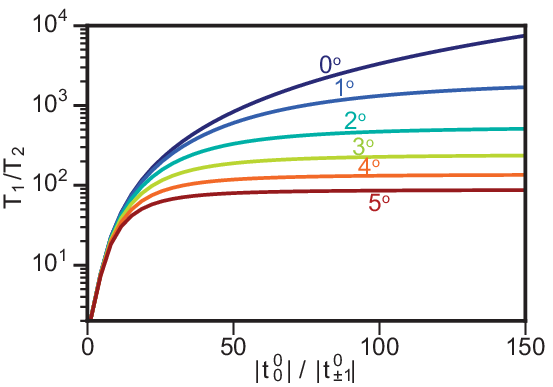}
\caption{Plot of $T_{1,a}/T_{2,0\leftrightarrow\pm1}$ vs. $\mid t_0^0\mid/\mid t_{\pm1}^0\mid$ from Eq. \ref{eq:T1T2Rat} for several values of $\theta$.}
\label{fig:T1T2RatioOrientation} 
\end{figure}

The singlet model predicts that the observed ratio of $T_1$ to $T_2$ in Figs. \ref{fig:T1T2Temp} and \ref{fig:T1Orientation} is strongly dependent on the ratio of the overlap parameters at zero field.  The analytical form of this dependence is shown in Eq. \ref{eq:T1T2Rat} which is plotted in Fig. \ref{fig:T1T2RatioOrientation}.  This figure shows that this ratio is strongly dependent on the orientation of the magnetic field, indicating that the best way to extract the ratio of the zero field overlap parameters is by performing a global fit across all orientations (Fig. \ref{fig:T1Orientation}).

\subsection{Model for spin relaxation: Orbach process with a singlet excited state}

Here we present a detailed analytical derivation of the spin relaxation of SiV$^0$ for an Orbach process mediated by a spin singlet excited state. The neutral silicon vacancy center has $D_{3d}$ symmetry with a ground spin-triplet state ($^3A_{2g}$), and the first excited singlet state is expected to be $^1E_g$.  The splitting between these states is unknown.  At zero magnetic field the triplet and singlet states can mix through spin-orbit coupling assisted by phonons \cite{Goldman2015PRL, 2015GoldmanPRB,2017Thiering,2011Doherty}:

\begin{equation}
\begin{split}
&\mid ^3\!\!\bar{A}_{2g}^{m_s=0} \rangle = \mid ^3\!\!A_{2g}^{m_s=0}\rangle +t_0^0 \mid ^1\!\!E_g \rangle\\
&\mid ^3\!\!\bar{A}_{2g}^{m_s=+1} \rangle = \mid ^3\!\!A_{2g}^{m_s=+1}\rangle +t_{+1}^0 \mid ^1\!\!E_g \rangle\\
&\mid ^3\!\!\bar{A}_{2g}^{m_s=-1} \rangle = \mid ^3\!\!A_{2g}^{m_s=-1}\rangle +t_{-1}^0 \mid ^1\!\!E_g \rangle\\
&\mid ^1\!\!\bar{E}_g \rangle = \mid ^1\!\!E_g \rangle+ \sum_{m_s} t_{m_s}^0\mid ^3\!\!A_{2g}^{m_s}\rangle\text{,}
\end{split}
\end{equation}
where $t_{m_s}^0$ are state mixing coefficients.  In the main text we refer to them as overlap parameters that connect the singlet and triplet subspaces since $t_{m_s}^0=\langle ^3\!\!\bar{A}_{2g}^{m_s} \mid \!\!^1\!\bar{E}_g \rangle$.  The $t_{m_s}^0$ coefficients arise from spin-orbit coupling and thus depend only on the orbital symmetry of the involved zero-field states, which is independent of the applied magnetic field.

The triplet eigenstates in the presence of a magnetic field can be found using a Wigner rotation to transform the eigenstates of the zero field splitting term from the molecular frame to the laboratory frame (the frame in which the Zeeman interaction is diagonal).  This model assumes that in a magnetic field the eigenstates of the spin Hamiltonian have mostly Zeeman character and the zero field splitting term can be neglected ($g\mu_B B/h \gg D$).  A general rotation, $R$, can be expressed in terms of Euler angles:

\begin{equation}
R\left(\alpha, \beta, \gamma\right) = R_{\hat{z}}(\gamma)R_{\hat{n}}(\beta)R_{\hat{z}}(\alpha)\text{,}
\end{equation}
where $\vec{\Omega}=(\alpha,\beta,\gamma)$ is the set of Euler angles following the ``passive" convention.  Under this rotation the irreducible tensors in the spin Hamiltonian $T_{J,m}$ transform to $\rho_{J,m}$ as:

\begin{equation}
\begin{split}
\rho_{J,m} = & R(\alpha,\beta,\gamma)T_{J,m}R^{-1}(\alpha,\beta,\gamma)=\\
& \sum_{m'} D^J_{m',m}(\alpha,\beta,\gamma)T_{J,m'}\text{,}
\end{split}
\end{equation}
where $D^{J}_{m',m}(\Omega)$ is the Wigner matrix of rank $J$.  The elements of this matrix are:
\begin{equation}
D^J_{m',m}(\alpha,\beta,\gamma) = \exp(-i m' \alpha)d^J_{m',m}\left( \beta \right) \exp(-i m \gamma)\text{,}
\end{equation}
with
\begin{equation}
d^J_{m',m}\left( \beta \right) = \int_{\theta,\phi}\!d\Omega~~Y_{Jm'}^*\left(\theta,\phi\right)e^{-\frac{i}{\hbar}\beta J_{\hat{n}}} Y_{Jm}\left( \theta, \phi \right)\text{,}
\end{equation}
where $Y_{Jm'}\left(\theta,\phi\right)$ are the standard spherical harmonic functions and $J_{\hat{n}}$ is the component of the total angular momentum along $\hat{n}\parallel \langle 1 1 0\rangle$.  Then for $J=S=1$:

$D^{S=1}_{m',m}\left(\alpha,\beta,\gamma\right)=$
\begin{equation}
  \begin{pmatrix}
     \frac{1+\cos(\beta)}{2}e^{-i\left(\alpha+\gamma\right)}&-\frac{1}{\sqrt{2}}\sin(\beta)e^{-i\alpha}& \frac{1-\cos(\beta)}{2}e^{-i\left(\alpha-\gamma\right)}\\
     \frac{1}{\sqrt{2}}\sin(\beta)e^{-i\gamma}&\cos(\beta)&-\frac{1}{\sqrt{2}}\sin(\beta)e^{i\gamma}\\
   \frac{1-\cos(\beta)}{2}e^{i\left(\alpha-\gamma\right)}&\frac{1}{\sqrt{2}}\sin(\beta)e^{i\alpha}& \frac{1+\cos(\beta)}{2}e^{i\left(\alpha+\gamma\right)}
  \end{pmatrix}\text{.}
\end{equation}
If we specifically define $R$ as the rotation away from $\langle 111 \rangle$ about the $\langle 110 \rangle$ axis, so that $\alpha =\varphi$, $\beta=\theta$, and $\gamma=0^\circ$ define the orientation of the magnetic field, the mixing of the transition amplitudes is given by:

\begin{equation}
t_{m'}= \sum_m D^{S=1}_{m',m}\left(\varphi,\theta,0\right) t_{m}^0\text{,}
\label{eq:rateTransform}
\end{equation}
From this we obtain the transition rates ($\left|t_{m}\right|^2$) by invoking the random phase approximation to neglect the cross terms (averaging over $\varphi$).  The physical origin of the random phase approximation can arise from taking an ensemble average over a bath of phonons that randomly induce transitions to the excited state through spin-orbit coupling.  The result is:
\begin{equation}
\left|t_{m'}\right|^2= \sum_m \langle\left|D^{S=1}_{m',m}\left(\varphi,\theta,0\right)\right|^2\rangle_\varphi \left|t_{m}^0\right|^2\text{,}
\label{eq:rateTransform}
\end{equation}
where
$\langle\left|D^{S=1}_{m',m}\left(\varphi,\theta,0\right)\right|^2\rangle_\varphi=$
\begin{equation}
  \begin{pmatrix}
     \cos^4(\theta/2)&\frac{1}{2}\sin^2(\theta)& \sin^4(\theta/2)\\
     \frac{1}{2}\sin^2(\theta)&\cos^2(\theta)&\frac{1}{2}\sin^2(\theta)\\
   \sin^4(\theta/2)&\frac{1}{2}\sin^2(\theta)& \cos^4(\theta/2)
    
  \end{pmatrix}\text{.}
\label{eq:MixMat}
\end{equation}
In the main text, Eqns. \ref{eq:T1General} for the overlap coefficients in the presence of an off-axis magnetic field are obtained by substituting Eq. \ref{eq:MixMat} into Eq. \ref{eq:rateTransform}.  

Next, the transition rate matrix (Eq. \ref{eq:MixMat}) can be used to model the spin relaxation processes for $S=1$ where the populations $\bm{P}=\left(P_{-1},P_0,P_{+1}\right)$ evolve according to:

\begin{equation}
\frac{d\bm{P}(t)}{dt}= \tilde{R}\bm{P}(t)\text{,}
\end{equation}
where the rate matrix $\tilde{R}$ is given by:

\begin{equation}
\begin{aligned}
\tilde{R}_{m,m'}=&C\left(1-\delta_{m,m'}\right)\mu^{m-m'}\left|t_m\right|^2\left|t_{m'}\right|^2-\\
&C\delta_{m,m'}\left(\sum_{m'' \neq m}\left|t_{m}\right|^2\left|t_{m''}\right|^2\mu^{m''-m}\right)
\end{aligned}\text{,}
\end{equation}
where $\delta_{m,m'}$ is the Kronecker delta function and $\mu=\exp(hf/kT)$ is the Boltzmann factor at $T = 30$ K and $f = 9.7$ GHz.  Assuming that $\left|t_{+1}^0\right| = \left|t_{-1}^0\right|$ and $\mu=1$, this results in two distinct rate eigenvalues $\lambda_1$, $\lambda_2$ corresponding to $T_{1,a}=\lambda_1^{-1}e^{-E_a/kT}$ and $T_{1,b}=\lambda_2^{-1}e^{-E_a/kT}$:

\begin{equation}
\begin{aligned}
&T_{1,a}=\frac{2e^{-E_a/kT}}{3C\left|t_0\right|^2\left(\left|t_{+1}\right|^2+\left|t_{-1}\right|^2\right)}\\
&T_{1,b}=\frac{e^{-E_a/kT}}{C\left|t_{-1}\right|^2\left(2\left|t_{+1}\right|^2+\left|t_0\right|^2\right)}\text{.}
\end{aligned}
\label{eq:eigenRate}
\end{equation}
Eqns. \ref{eq:eigenRate} were used to simulate the angular dependence of $T_1$ as a function of $\left|t_0\right|/\left|t_{\pm1}\right|$ in Figs. \ref{fig:T1Orientation}a and \ref{fig:PJTEModel}b.  If we assume that $\left|t_0^0\right|\gg \left|t_{\pm1}^0\right|$  then Eqns. \ref{eq:eigenRate} reduce to Eqns. \ref{eq:T1e_angle}.

\subsection{Model for spin relaxation: Orbach process with a triplet excited state}
The excited state can also be a spin triplet state, such as a quasilocalized vibronic mode or a low lying electronic state.  For this model we define two S=1 spin Hamiltonians for the ground state ($\hat{H}_g$) and excited state ($\hat{H}_e$) that differ only in their zero field splitting tensors ($\tilde{D}_g\neq \tilde{D}_e$):

\begin{equation}
\begin{split}
&\hat{H}_g=\hat{\bm{S}}^\dagger\tilde{D}_g\hat{\bm{S}} + \mu_B \hat{\bm{S}}^\dagger\tilde{g}\bm{B}\\
&\hat{H}_e=\hat{\bm{S}}^\dagger\tilde{D}_e\hat{\bm{S}} + \mu_B \hat{\bm{S}}^\dagger\tilde{g}\bm{B}\text{,}
\end{split}
\end{equation}
with eigenstates $\mid \!\!m_s\rangle_g$ and $\mid \!\!n_s\rangle_e$, respectively.  The rate matrix describing the spin relaxation is given by:

\begin{equation}
\begin{aligned}
&R_{m.m'}=C\left(1-\delta_{m,m'}\right)\sum_{n}\left|_g\langle m \mid n\rangle _{e~e} \langle n \mid m' \rangle_g\right|^2 \mu^{m-m'}-\\
&C\delta_{m,m'}\left(\sum_{m'' \neq m}\mu^{m''-m}\sum_{n}\left|_g\langle m'' \mid n\rangle_{e~e} \langle n \mid m \rangle_g \right|^2\right)\text{.}
\end{aligned}
\label{eq:eigenRateTrip}
\end{equation}

\begin{figure}
\includegraphics[width=\linewidth]{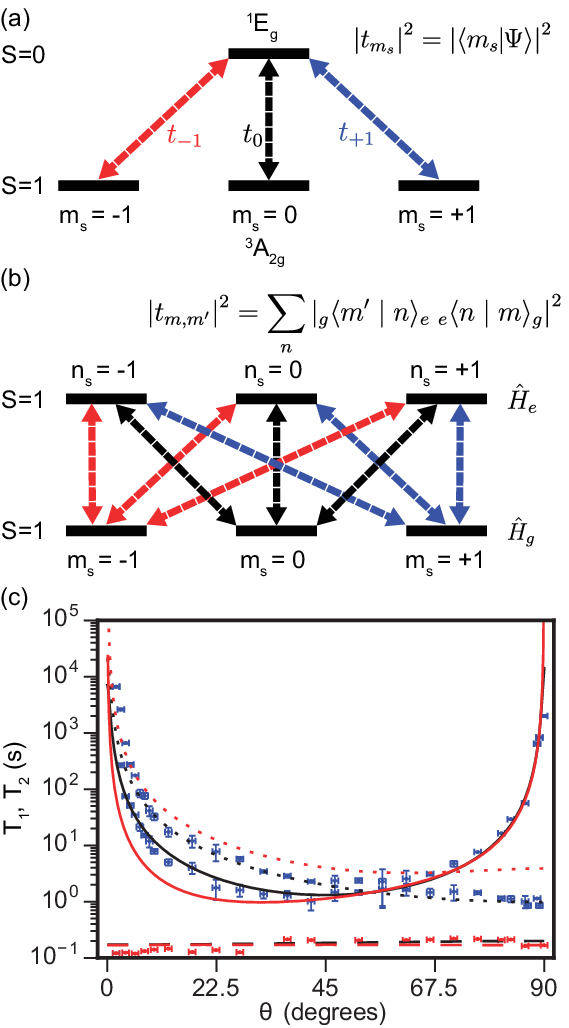}
\caption{(a) Model for the Orbach process of SiV$^0$ with a singlet excited state.  The transition rates ($\left|t_{m_s}\right|^2$) are determined by spin orbit coupling and depend on the overlap between the electronic wavefunctions of the ground triplet state and the excited singlet state. (b) Model for the Orbach process of SiV$^0$ with a triplet excited state.  The transition rates between the ground state spin sublevels depend on the overlap between the spin eigenstates of the ground state and excited state, and must be summed over all spin sublevels in the excited state. (c) $T_1$ and $T_2$ orientation dependence from Fig. \ref{fig:T1Orientation}a, plotted against calculated fits from the singlet (black) and triplet (red) Orbach models: (solid line) $T_{1,a}$, (dashed line) $T_{1,b}$, and (long dash) $T_2$.  The singlet model fit assumes $\left|t_0^0/t_{\pm1}^0\right|=125$.  The triplet model fit assumes a coaxial excited state ZFS tensor with $D_e=5$ GHz.}
\label{fig:STCompare} 
\end{figure}

In the triplet model spin flips can occur through any of the three spin sublevels of the excited state (Fig. \ref{fig:STCompare}b), increasing the complexity of the rate matrix.  Spin relaxation arises from the overlap between the eigenstates of the two triplet states, and slight variations in the character of the states become important.  Thus the zero field splitting terms for both the ground state and excited state cannot be neglected when calculating the triplet state overlap coefficients and the rate matrix ($R_{m,m'}$).  For these reasons the analytical solution for the Orbach model with a triplet excited state is not compact, and instead we numerically simulate the spin relaxation by diagonalizing the rate matrix Eq. \ref{eq:eigenRateTrip}. 

In general $\tilde{D}_e$ can differ from $\tilde{D}_g$ in either its quantization axis, magnitude of the axial component, or magnitude of the rhombicity parameter.  In the case where the quantization axis of the excited state is not aligned with the quantization axis of the ground state (e.g. due to an E type quasilocalized vibronic mode that breaks $D_{3d}$ symmetry) the resulting orientation dependence qualitatively disagrees with the $T_1$ data.  The same disagreement was found to be true for the case where rhombicity was introduced into the excited state spin Hamiltonian. However, the orientation dependence of $T_1$ can be partially reproduced by assuming that $\tilde{D}_e$ is axial (no rhombicity) and also coaxial with the ground state $\tilde{D}_g$, thus preserving $D_{3d}$ symmetry.  Focusing on just fitting the $T_1$ orientation dependence (ignoring $T_2$), the closest fit to the $T_1$ data was found with $D_e=1$ GHz.  However, this excited state zero field splitting tensor predicts that $T_2\sim 100$ ns, which is inconsistent with the measured values. As described in the main text, the Zeeman energy is large, and in order to reproduce the $T_2$ data, $D_e$ needs to be comparable to the Zeeman energy, and the $T_2$ data is reproduced best by the triplet model when $D_e\approx 5-7$ GHz (Fig. \ref{fig:T2Orientation}). The simulated orientation dependence of $T_1$ for this excited state zero field splitting is qualitatively similar to the data, but lies outside of the error bars for both time constants (Fig. \ref{fig:STCompare}c). Furthermore, such a large difference in zero field splitting between the ground and excited states is unlikely.

\subsection{Orientation dependence of $T_2$}

\begin{figure}
\includegraphics[width=\linewidth]{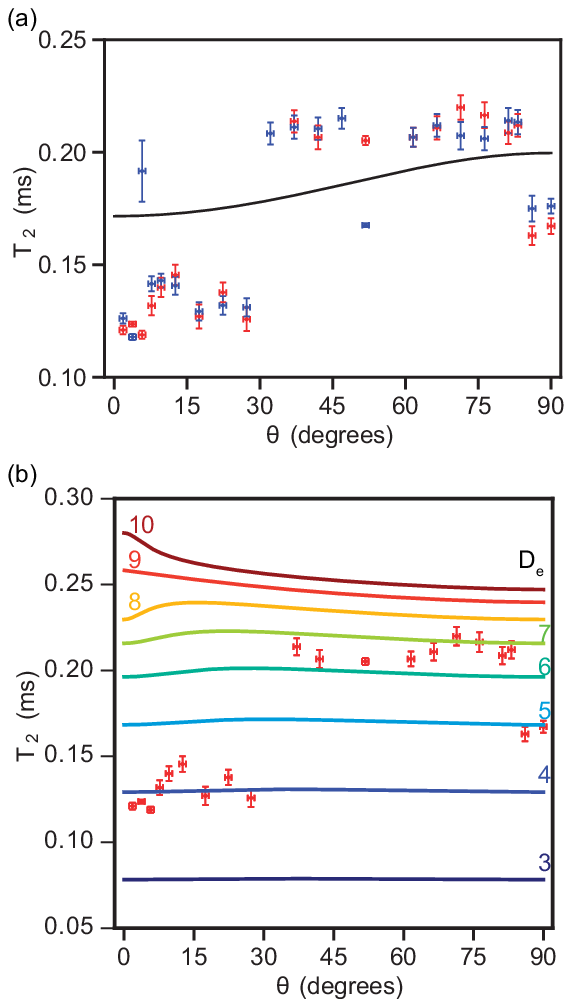}
\caption{Orientation dependence of the SiV$^0$ electron spin coherence time, $T_2$ (replotted from Fig. \ref{fig:T1Orientation}a).  (a) Predicted orientation dependence using an Orbach model with a singlet excited state (black curve, Eq. \ref{eq:T2Orien}) with no free parameters.  Measurements were made on both the $m_s = 0 \leftrightarrow +1$ transition (red points) and the $m_s = -1 \leftrightarrow 0$ transition (blue points) at each orientation.  (b) Predicted orientation dependence of an Orbach model with a triplet excited state.  The simulated curves are shown for several values of $D_e$ (labeled in units of GHz).}
\label{fig:T2Orientation} 
\end{figure}

Our model for the Orbach process predicts a weak orientation dependence of $T_2$.  The orientation dependence fits shown in Fig.~\ref{fig:T1Orientation}a utilize the same overlap amplitudes ($t_0^0$, $t_{\pm1}^0$) to explain both $T_1$ and $T_2$.  The actual expression used in fitting the $T_2$ dependence in Fig. \ref{fig:T1Orientation}a is given by:

\begin{equation}
\begin{split}
\frac{1}{T_{2,0\leftrightarrow \pm1}}&=\frac{1}{3}C\left(\left|t_0^0\right|^2+2\left|t_{\pm1}^0\right|^2\right)\left(\left|t_{0}\right|^2+\left|t_{\pm1}\right|^2\right)+\\
&~~~~~~~~~~~\frac{1}{T_{2(ID)}}+\frac{1}{T_{2(SD)}}\text{,}
\end{split}
\label{eq:T2Orien}
\end{equation}
which in addition to the Orbach process also includes instantaneous diffusion and $^{13}$C spectral diffusion mechanisms.  We used $T_{2(ID)}=0.319$ ms and $T_{2(SD)}=0.95$ ms in these simulations.  

The orientation dependence of $T_2$ is shown in Fig. \ref{fig:T2Orientation} with the simulated fits according to the singlet (Fig. \ref{fig:T2Orientation}a) and triplet (Fig. \ref{fig:T2Orientation}b) models, using the $t_{m_s}^0$ and $C$ parameters determined from the $T_1$ data.  The singlet model has no other free fitting parameters, and we plot Eq. \ref{eq:T2Orien} for the singlet model assuming that $\left|t^0_0/t_{\pm1}^0\right|=125$ as determined from the fit of the $T_1$ orientation dependence (Fig. \ref{fig:T1Orientation}a).  The singlet model predicts the magnitude of $T_2$ with reasonable accuracy.

The triplet model has four free parameters, two angles that set the quantization axis of the excited state, the axial part of the zero field splitting tensor, and the rhombic part of the zero field splitting tensor.  We only consider the case where the zero field splitting tensor of the excited state is axial and aligned with the symmetry axis of the defect since this is the case that best produces the measured $T_1$ orientation dependence (Fig. \ref{fig:STCompare}c).  The dependence for several values of $D_e$ is shown and the best fit occurs with $D_e\approx 5-7$ GHz. Alternatively, if the spin does not fully decohere through a single cycle through the excited state, the magnitude of $T_2$ can be larger than the simulated values.






\end{document}